\documentclass[journal,11pt,draftcls,peerreview,onecolumn]{IEEEtran}

\usepackage{cite}      

\usepackage{graphicx}  

\usepackage{amssymb}

\newcounter{mytempeqncnt}

\begin{document}
%
\title{Nonregenerative MIMO Relaying with Optimal Transmit Antenna Selection}
%
%
\author{Steven~W.~Peters,~\IEEEmembership{Student Member,~IEEE,} and~Robert~W.~Heath,~Jr.,~\IEEEmembership{Senior~Member,~IEEE}
\thanks{The authors are with the Wireless Networking and Communications Group, Department of Electrical and Computer Engineering, 
1 University Station C0803, University of Texas at Austin, 
Austin, TX, 78712-0240 (email: \{peters, rheath\}@ece.utexas.edu, phone: (512) 471-1190, fax: (512) 471-6512). EDICS: COM-\{ESTI,MIMO,NETW\}}
\thanks{This work was supported by the Semiconductor Research Corporation under contract 2007-HJ-1648.}}%

\markboth{Accepted to IEEE Signal Processing Letters, January~2008}%
{Peters and Heath: Nonregenerative MIMO Relaying with Optimal Transmit Antenna Selection}

\setcounter{page}{1}
\maketitle

\begin{abstract}
We derive optimal SNR-based transmit antenna selection rules at the source and relay for the nonregenerative half duplex MIMO 
relay channel.  While antenna selection is a suboptimal form of beamforming, it has the advantage that the optimization is 
tractable and can be implemented with only a few bits of feedback from the destination to the source and relay. 
We compare the bit error rate of optimal antenna selection at both the 
source and relay to other proposed beamforming techniques and propose methods for performing the necessary limited feedback.
\end{abstract}



\newcommand{\sr}[1]{$\mathcal{#1}$}
\newcommand{\vecnorm}[2]{\left|\tilde{\bf H}_{#1}^{#2}\right|}

\section{Introduction}
\PARstart{D}{espite} the lack of precise knowledge of its basic theoretical behavior and limits, relaying is beginning
to find practical application in standards such as IEEE 802.16j~\cite{IEEE80216j}. By deploying relatively inexpensive relays, service providers 
can reduce the number of base stations required to serve a given area, or increase capacity at the cell edge.

Relaying research efforts have also increased recently~\cite{Wang2005,Lo2005,Munoz-Medina2007,Fan2007a,Yuksel2007,Tang2007}.
Capacity bounds for the full-duplex MIMO relay channel were derived in~\cite{Wang2005,Lo2005}.
The authors of~\cite{Yuksel2007} derive the optimal infinite-SNR diversity-multiplexing tradeoff for the half duplex MIMO relay channel and 
find that a compress-and-forward strategy is optimal in this sense. Recently, practical strategies have been developed for MIMO relaying.
Both~\cite{Munoz-Medina2007} and~\cite{Tang2007} derive the mutual-information-maximizing nonregenerative linear relay for spatial multiplexing 
when the direct link is ignored. 


This letter derives the optimal transmit antenna selection criteria at both source and relay; i.e., all transmissions occur using the transmit
antenna that will give the destination the highest post-processing signal-to-noise ratio. We consider the case where only a single spatial stream
is to be sent from source to destination. This scenario arises when the channel is ill-conditioned (i.e., there is a dominant path of propagation
in the source-destination channel), or if robustness via diversity is preferred over throughput (i.e., near the cell edge). 

Unlike most previous practial MIMO relay results (e.g.,~\cite{Munoz-Medina2007,Fan2007a,Tang2007}), the strategy derived here is the optimal 
transmit antenna selection strategy when the direct link from source to relay is not ignored.
We prove that transmit antenna selection, combined with an MMSE receiver at the destination, achieves the full diversity order of the 
MIMO single relay channel. That is, at high SNR the probability of outage decays with SNR as quickly as is possible in such a model. 
Further, antenna selection requires less feedback than beamforming. Distributed space-time codes, which may also achieve the full diversity
gain, not only require their own level of overhead for coordination and synchronization, but also require the relay to be able to decode the 
message transmitted by the source.

Compared to recent results using limited feedback beamforming~\cite{Khoshnevis2008}, under the tested parameters given
in the aforementioned paper, antenna selection at both source and destination is about 
twice as likely to cause bit errors as a Grassmannian codebook with 16 codes, which is a loss of about 1 dB at high SNR. 
In return, antenna selection requires only $\log_2 N_SN_R$ bits of feedback versus $3\log_2 N+2b$ bits in~\cite{Khoshnevis2008},
where $N_S$ and $N_R$ are the number of antennas at the source and relay, respectively, $N$ is the size of the Grassmannian codebook, and
$b$ is the quantization in bits of the SNR feedback required in~\cite{Khoshnevis2008}.

This letter uses capital boldface letters to refer to matrices and lowercase boldface letters for column vectors.
The notation $\|{\bf h}\|$ refers to the L2-norm of the vector $\bf h$, and ${\bf H}^*$ is the complex conjugate transpose of the matrix $\bf H$.
The vector ${\bf h}^{(i)}$ refers to the $i$th column of the matrix $\bf H$. Finally, 
\begin{eqnarray}
A \doteq B \iff \lim_{\texttt{SNR}\rightarrow\infty}\frac{\log A}{\log\texttt{SNR}} = -B.\nonumber
\end{eqnarray}

\section{System Model \& Antenna Selection}
We assume a single source \sr{S} transmitting information to a destination \sr{D} with a single relay \sr{R} aiding the transmission. 
The source, destination, and relay are equipped with $N_S$, $N_D$, and $N_R$ antennas, respectively. All nodes
operate in half-duplex mode. Unlike most prior work in MIMO relaying, \emph{we do not ignore the direct link between
\sr{S} and \sr{D}}. 

The source \sr{S} wishes to transmit the scalar symbol
$s$ to \sr{D}, where $\mathbb{E}|s|^2=E_s=\texttt{SNR}$, $E_s$ is the average power constraint at both \sr{S} and \sr{R}, and 
$\sigma^2=1$ is the overall noise power at each node. Since the signal-to-noise 
ratio is the metric of interest, an imbalance of noise energy among the nodes can be modeled in the appropriate fading parameter 
for ${\bf H}_{XY}$. 
For instance, if the relay has noise power $\sigma_r^2$, in an independent Rayleigh fading 
environment these definitions would change the channel fading parameter of the corresponding exponential distribution from $\lambda_{SR}$ to 
$\lambda_{SR}\sigma_r^2$.

We denote the channel from \sr{X} to \sr{Y}, $\mathcal{X}\in\{\mathcal{S},\mathcal{R}\}$, 
$\mathcal{Y}\in\{\mathcal{R},\mathcal{D}\}$, $\mathcal{X}\ne\mathcal{Y}$, as ${\bf H}_{XY}$, 
and ${\bf h}_{XY}^{(i)}$ is the vector channel from the $i$th transmit antenna at \sr{X} to \sr{Y}. We also define 
\begin{equation}
\gamma_{XY}^{(i)}=\left\|{\bf h}_{XY}^{(i)}\right\|^2\texttt{SNR}
\end{equation}
to be the equivalent receive SNR from $\mathcal{X}_i\to\mathcal{Y}$.

We assume the block fading model. In the first stage, if \sr{S} transmits $s$ on antenna $i$, \sr{R} receives the signal
\begin{equation}
{\bf y}_R = {\bf h}_{SR}^{(i)}s+{\bf n}_R,
\end{equation}
where ${\bf n}_R$ is the zero-mean spatially white complex Gaussian noise vector with covariance $\sigma^2{\bf I}_{N_R}$ as observed by 
\sr{R}. Since the relay is also transmitting on only one of its antennas, it must combine its received vector to form a single symbol. 
It can be shown that the optimal way to do this is to perform MRC on the signal, resulting in a scalar 
\begin{equation}
s_R=\alpha({\bf h}_{SR}^{(i)})^*{\bf y}_R,
\end{equation}
where $\alpha$ is the scaling factor to ensure \sr{R} transmits at its expected power constraint; i.e.,
\begin{equation}
\alpha^2=\frac{1}{\|{\bf h}_{SR}^{(i)}\|^4+\|{\bf h}_{SR}^{(i)}\|^2/\texttt{SNR}}.
\end{equation}
At \sr{D}, the first stage results in
\begin{equation}
{\bf y}_{D,1} = {\bf h}_{SD}^{(i)}s+{\bf n}_{D,1}.
\end{equation}
In the second stage, \sr{R} transmits $s_R$ to \sr{D} on antenna $k$:
\begin{equation}
{\bf y}_{D,2} = {\bf h}_{RD}^{(k)}s_R+{\bf n}_{D,2}.
\end{equation}
The destination now has two observations containing $s$. To put the channel in standard MIMO notation, we define 
\begin{eqnarray}
	{\bf h} & = & \left(
		\begin{array}{c}
			{\bf h}_{SD}^{(i)}\\
			\frac{\|{\bf h}_{SR}^{(i)}\|{\bf h}_{RD}^{(k)}}{\sqrt{\|{\bf h}_{SR}^{(i)}\|^2+1/\texttt{SNR}}}
		\end{array}\right)\\
	{\bf n} & = & \left(
		\begin{array}{c}
			{\bf n}_{D,1}\\
			\frac{{\bf h}_{RD}^{(k)}({\bf h}_{SR}^{(i)})^*{\bf n}_{R}}{\|{\bf h}_{SR}^{(i)}\|\sqrt{\|{\bf h}_{SR}^{(i)}\|^2+1/\texttt{SNR}}}
				+{\bf n}_{D,2}
		\end{array}\right)\\
	{\bf y}_D & = & \left(\begin{array}{c}{\bf y}_{D,1}\\{\bf y}_{D,2}\end{array}\right)
\end{eqnarray}
so that
\begin{equation}
	{\bf y}_D = {\bf h}s+{\bf n}.
\end{equation}
\begin{figure*}[!t]
\normalsize
\setcounter{mytempeqncnt}{\value{equation}}
\setcounter{equation}{10}
\begin{equation}
\label{eqn:SNR_i}
\gamma^{(i)}=\gamma_{SD}^{(i)}\left(
				\frac{\gamma_{SD}^{(i)}(\gamma_{SR}^{(i)}+1)^2+\gamma_{SR}^{(i)}\gamma_{RD}^{(k)}(
				\gamma_{SR}^{(i)}+1+\gamma_{SR}^{(i)}+1+\gamma_{SR}^{(i)}\gamma_{RD}^{(k)})}
				{\gamma_{SD}^{(i)}(\gamma_{SR}^{(i)}+1)^2+\gamma_{SR}^{(i)}\gamma_{RD}^{(k)}(\gamma_{SR}^{(i)}+1+
				\gamma_{RD}^{(k)})}
			\right)
\end{equation}
\setcounter{equation}{\value{mytempeqncnt}}
\hrulefill
\vspace*{4pt}
\end{figure*}
\setcounter{equation}{11}
We assume the destination \sr{D} now applies a linear filter $\bf w$ to ${\bf y}_D$ to obtain an estimate of $s$. Although suboptimal, we will see
later that in some cases the destination may wish to apply MRC (${\bf w}={\bf h}$) on ${\bf y}_D$, and doing so would result in the post-processing 
signal-to-noise 
ratio $\gamma^{(i)}$ of (\ref{eqn:SNR_i}) at the top of the page. In this form, it is easy to see that, if 
\begin{equation}
	\left\{\gamma_{SR}^{(i)}<\gamma_{SD}^{(i)}\right\}\bigcap
		\left\{\gamma_{RD}^{(k)}>\gamma_{SD}^{(i)}(\gamma_{SR}^{(i)}+1)/(\gamma_{SD}^{(i)}-\gamma_{SR}^{(i)})\right\},
\end{equation}
then $\gamma^{(i)}<\gamma_{SD}^{(i)}$ and relaying is worsening performance. This occurs when the SNR from
\sr{R} to \sr{D} is very good relative to the others, and the SNR from \sr{S} to \sr{R} is worse than the direct SNR. Effectively,
the \sr{R} to \sr{D} channel is dominating the received signal, but it consists of mostly noise relative to the direct signal. Recall that
MRC is only optimal when the observations contain the same noise variance~\cite{Goldsmith2005}. 
Because of the amplified noise at \sr{R}, this is
not the case here. In this case, one can show that the optimal receive filter in the minimum mean-squared error (MMSE) sense is
\begin{equation}
	{\bf w} = {\bf R}_{{\bf y}_D}^{-1}{\bf R}_{{\bf y}_Ds},
\end{equation}
where ${\bf R}_{{\bf y}_D}=\mathbb{E}\{{\bf y}_D{\bf y}_D^*\}$ and ${\bf R}_{{\bf y}_Ds}=\mathbb{E}\{{\bf y}_Ds^*\}$. The post-processing
SNR is then
\begin{equation}
	\label{eqn:easy_SNR}
	\gamma^{(i)}=\gamma_{SD}^{(i)} + \frac{\gamma_{SR}^{(i)}\gamma_{RD}^{(k)}}{\gamma_{SR}^{(i)}+\gamma_{RD}^{(k)}+1}.
\end{equation}
Note that this requires the destination to 
have knowledge of $\|{\bf h}_{SR}\|$. If this is not possible, suboptimal MRC resulting in the SNR of
(\ref{eqn:SNR_i}) may be used instead, which requires less training. A method for obtaining this CSI is presented in Section~\ref{sec:feedback}.

Note that in (\ref{eqn:easy_SNR}), for fixed $\gamma_{SD}^{(i)}$ and $\gamma_{SR}^{(i)}$, $\gamma^{(i)}$ is maximized when $\gamma_{RD}^{(k)}$
is maximized. Thus, the antenna selection at the relay is independent of the selection at the source, and we can substitute the index of the 
optimal relay transmit antenna $k_o$ in for $k$ in all subsequent equations. The same cannot be said of the regular MRC equation (\ref{eqn:SNR_i}).

Finally, we note that antenna selection at the relay is suboptimal, and the optimal strategy in this case is intuitive; since the SNR
expression (\ref{eqn:easy_SNR}) is the addition of the independent SNR terms for the parallel channels to the destination from the source,
the relay should apply a filter that maximizes the SNR to the destination. One can show that
this filter is
${\bf W}={\bf v}^{(1)}({\bf h}_{SR}^{(i_o)})^*$, where ${\bf v}^{(1)}$ is the right singular vector of ${\bf H}_{RD}$ corresponding to its largest
singular value, and $i_o$ is the index of the source antenna that maximizes (\ref{eqn:easy_SNR}). Intuitively, ${\bf W}$ is the combination of a
receive filter matched to ${\bf H}_{SR}$ and a transmit beamforming vector matched to ${\bf H}_{RD}$. Implementing this filter would require 
perfect knowledge of ${\bf H}_{RD}$ at the relay and an SVD operation. All of our results hold with
this optimal strategy, with $\|{\bf h}_{RD}^{(k_o)}\|^2$ replaced with $\lambda_{RD}=\sigma_{RD}^2$, the square of the largest singular
value of ${\bf H}_{RD}$.

\section{Training and Limited Feedback}
\label{sec:feedback}
We now discuss how channel state information might be obtained in the channel of interest so that a reliable antenna selection
strategy may be implemented. All three channels need to be estimated at their respective receivers; this can be accomplished using previously
studied MIMO training methods. Only knowledge of the link SNRs (i.e., $\gamma_{XY}^{(i)}$'s) is required for transmit antenna selection. 
Therefore a low complexity signal,
such as a short narrowband tone, may be used for estimating SNR to choose an antenna to train from. 
This is first sent from \sr{R} to \sr{D} from each relay antenna. \sr{D} then feeds back which antenna \sr{R} should
use to transmit, and, from this antenna, a training sequence suitable for channel estimation is sent to the destination. The source repeats this
process with its transmit antennas, with the relay forwarding its received signal on its optimal antenna. This way, the destination can estimate
the SNR between the source and relay to perform MMSE combination as described earlier. 

The destination finds (\ref{eqn:easy_SNR}) for each source antenna, then feeds back to the source the index of the antenna that resulted in 
the largest $\gamma^{(i)}$. The source then transmits a training sequence from this antenna, which does not need to be forwarded by the 
relay. This process requires $\log(N_SN_D)$ bits of feedback, two time slots of training, and $N_R+2N_S$ time slots for SNR estimation. 
Minimizing the time required for SNR estimation is thus important for this feedback strategy.

\section{Diversity Analysis}
Antenna selection is used to exploit the diversity gain available in the channel. Using~(\ref{eqn:easy_SNR}) we now show
that this strategy achieves full diversity gain. We first give an upper bound on the diversity order of the half-duplex MIMO relay channel
when the source and destination transmit orthogonally in equal time slots. Yuksel and Erkip~\cite{Yuksel2007} have derived this result 
for arbitrary time sharing when the source is allowed to transmit in the second time slot, so this result is a special case of their derivation.
This derivation is included here to prove that our added restrictions (i.e., equal transmission times, source silent in the second time slot) do
not decrease the maximum diversity order of the channel.
We first define
\begin{eqnarray}
\label{eqn:IBC}
I_{BC} & = & I(S;Y_R,Y_{D,1})\\
\label{eqn:I1}
I_1 & = & I(S;Y_{D,1})\\
I_2 & = & I(S_R;Y_{D,2})\\
\label{eqn:IMAC}
I_{MAC} & = & I_1 + I_2,
\end{eqnarray}
where $S$ is the random variable corresponding to the transmitted signal from the source, $Y_R$ is the received signal at the relay, 
$Y_{D,n}$ is the received signal at the destination in the $n$th time slot, and $S_R$ is the transmitted signal at the relay.
Using equations (27) and (28) in~\cite{Yuksel2007} with $t=0.5$ and the source not transmitting in the second time slot, 
\begin{equation}
I(S;{\bf\it Y}_D)\le 0.5\min\{I_{BC},I_{MAC}\}.
\end{equation}
Now we can bound the probability of outage for a fixed $I_0$ as
\begin{eqnarray}
P_{out} & = & {\rm Pr}\{I(S;Y_D) < I_0\}\nonumber\\
{} & \ge & {\rm Pr}\left\{0.5\min\{I_{BC},I_{MAC}\}<I_0\right\}.
\end{eqnarray}
The event where the minimum of two variables is less than a constant is equivalent to the union of the events that each of the variables
is less than the constant. Defining $P_{out,BC}={\rm Pr}(I_{BC}<2I_0)$, and similarly for $P_{out,MAC}$, we can write
\begin{eqnarray}
P_{out} & \ge & {\rm Pr}\left(\{I_{BC}<2I_0\}\bigcup\{I_{MAC}<2I_0\}\right)\\
{} & = & P_{out,BC} + P_{out,MAC} -\nonumber\\
{} & {} & {\rm Pr}\left(\{I_{BC}<2I_0\}\bigcap\{I_{MAC}<2I_0\}\right).
\end{eqnarray}
Recall from (\ref{eqn:IMAC}) that $I_{MAC}$ is the sum of two nonnegative random variables. Such a sum is always less than or equal to twice the
maximum of the two random variables. Then, by making the codebook for $S_R$ independent from that of $S$, and defining 
$P_{out,I_1}={\rm Pr}(I_1<I_0)$ and $P_{out,I_2}$ similarly,
\begin{eqnarray}
P_{out} & \ge & P_{out,BC} + {\rm Pr}\{\max\{I_1,I_2\}<I_0\} -\nonumber\\
{} & {} & {\rm Pr}\left(\{I_{BC}<2I_0\}\bigcap\{I_{MAC}<2I_0\}\right)\\
{} & \ge & P_{out,BC} + P_{out,I_1}P_{out,I_2} -\nonumber\\ 
{} & {} & {\rm Pr}\left(\{I_{BC}<2I_0\}\bigcap\{I_{MAC}<2I_0\}\right).
\end{eqnarray}
Conversely, the sum of $I_1$ and $I_2$ is always greater than the maximum of the two. Also, note from (\ref{eqn:IBC}) and (\ref{eqn:I1}) that
$I_{BC}\ge I_1$ so that
\begin{eqnarray}
P_{out} & \ge & P_{out,BC} + P_{out,I_1}P_{out,I_2} -\nonumber\\ 
{} & {} & {\rm Pr}\left(\{I_{BC}<2I_0\}\bigcap\{\max\{I_1,I_2\}<2I_0\}\right)\nonumber\\
{} & = & P_{out,BC} + P_{out,I_1}P_{out,I_2} -\nonumber\\
{} & {} & {\rm Pr}\{\max\{I_{BC},I_2\}<2I_0\}.
\end{eqnarray}
Finally, again assuming independent channels on all links,
\begin{eqnarray}
P_{out} & \ge & P_{out,BC} + P_{out,I_1}P_{out,I_2} -\nonumber\\
\label{eqn:IBCindI2}
{} & {} & P_{out,BC}{\rm Pr}\{I_2<2I_0\}.
\end{eqnarray}
From MIMO information theory we know that (see~\cite{Yuksel2007}, Sec. III and IV)
\begin{eqnarray}
{\rm Pr}\{I_{BC}<c\} & \doteq & N_S(N_R+N_D)\\
{\rm Pr}\{I_1<c\} & \doteq & N_SN_D\\
{\rm Pr}\{I_2<c\} & \doteq & N_RN_D,
\end{eqnarray}
for all $c\in\mathbb{R}$. Thus, the last term in (\ref{eqn:IBCindI2}) will decay as $N_SN_R+N_SN_D+N_RN_D$ with $\log\texttt{SNR}$ and is 
thus irrelevant to the diversity analysis. The first term will decay as $N_S(N_R+N_D)$, while the second term decays as $N_D(N_S+N_R)$, so that
\begin{eqnarray}
\label{eqn:pout_upper}
P_{out} & \dot{\le} & N_SN_D + N_R\min\{N_S,N_D\}.
\end{eqnarray}

We now derive a lower bound on the diversity order of optimal antenna selection in flat i.i.d. Rayleigh fading by using (\ref{eqn:easy_SNR}).
First we define
\begin{eqnarray}
\label{eqn:SRD}
\gamma_{SRD}^{(i,k_o)}=\gamma_{SR}^{(i)}\gamma_{RD}^{(k_o)}/(\gamma_{SR}^{(i)}+\gamma_{RD}^{(k_o)}+1).
\end{eqnarray}
Since we choose the source transmit antenna that maximizes the SNR $\gamma$ at the destination,
\begin{eqnarray}
P_{out} & = & {\rm Pr}\{\gamma < \gamma_0\}\nonumber\\
{} & = & {\rm Pr}\{\max_i\{\gamma^{(i)}\} < \gamma_0\}\nonumber\\
{} & = & {\rm Pr}\{\max_i\{\gamma_{SD}^{(i)} + \gamma_{SRD}^{(i,k_o)}\}<\gamma_0\}.
\end{eqnarray}
As before, the sum of two random variables is greater than the maximum of the two. 
\begin{eqnarray}
P_{out} & \le & {\rm Pr}\{\max_i\{\max\{\gamma_{SD}^{(i)}, \gamma_{SRD}^{(i,k_o)}\}\} < \gamma_0\}\nonumber\\
{} & = & {\rm Pr}\{\max_i\{\gamma_{SD}^{(i)}, \gamma_{SRD}^{(i,k_o)}\} < \gamma_0\}.
\end{eqnarray}
Since each channel is mutually independent of the others, and the channel from each source antenna to the destination is also independent from
the others, we define $P_{out,SD}={\rm Pr}(\gamma_{SD}^{(1)}<\gamma_0)$, thus
\begin{eqnarray}
P_{out} & \le & {\rm Pr}\{\max_i\{\gamma_{SD}^{(i)}\} < \gamma_0\}{\rm Pr}\{\max_i\{\gamma_{SRD}^{(i,k_o)}\} < \gamma_0\}\nonumber\\
{} & = & \left(P_{out,SD}\right)^{N_S}{\rm Pr}\{\max_i\{\gamma_{SRD}^{(i,k_o)}\} < \gamma_0\}.
\end{eqnarray}
Now define 
\begin{equation}
\gamma_{M,i}=\min\{\gamma_{SR}^{(i)},\gamma_{RD}^{(k_o)}\}.
\end{equation} 
If $\gamma_{M,i}\ge1$, then 
$\gamma_{SRD}^{(i,k_o)}>\gamma_{M,i}/3$. Otherwise, $\gamma_{SRD}^{(i,k_o)}>(\gamma_{M,i})^2/3$. In either case, since
$\gamma_0$ is arbitrary, we let $\gamma_0>1/3$ and proceed\footnote{Since $P_{out}$ is monotone increasing with increasing $\gamma_0$, no loss
in generality occurs by assuming $\gamma_0>1/3$. For example, let $\gamma_L<1/3$. Then ${\rm Pr}(\gamma<\gamma_L)<{\rm Pr}(\gamma<\gamma_0)$. Thus,
if $P_{out}\doteq d$, then ${\rm Pr}(\gamma<\gamma_L)\dot{\ge} d$.}
\begin{eqnarray}
P_{out} & < & \left(P_{out,SD}\right)^{N_S}{\rm Pr}\{\max_i\{\gamma_{M,i}\}<3\gamma_0\}.
\end{eqnarray}
We can again split up the minimum event into a union:
\begin{eqnarray}
P_{out} & < & \left(P_{out,SD}\right)^{N_S}\times\nonumber\\
{} & {} & {\rm Pr}(\{\max_i\{\gamma_{SR}^{(i)}\}<3\gamma_0\}\bigcup\{\max_i\{\gamma_{RD}^{(k_o)}\}<3\gamma_0\})\nonumber\\
{} & = & \left(P_{out,SD}\right)^{N_S}\times\nonumber\\
{} & {} & \biggl[{\rm Pr}\{\max_i\{\gamma_{SR}^{(i)}\}<3\gamma_0\}+{\rm Pr}\{\gamma_{RD}^{(k_o)}<3\gamma_0\} -\nonumber\\
{} & {} & {\rm Pr}\{\max_i\{\gamma_{SR}^{(i)}\}<3\gamma_0\}{\rm Pr}\{\gamma_{RD}^{(k_o)}<3\gamma_0\}\biggr].
\end{eqnarray}
Again, since the channels between each source transmit antenna and the relay are independent, we define 
$P_{out,SR}={\rm Pr}(\gamma_{SR}^{(1)}<3\gamma_0)$ and $P_{out,RD}={\rm Pr}(\gamma_{RD}^{(k_o)}<3\gamma_0)$, and
\begin{eqnarray}
P_{out} & < & \left(P_{out,SD}\right)^{N_S}\times\nonumber\\
{} & {} & \biggl[\left(P_{out,SR}\right)^{N_S}+P_{out,RD} -\nonumber\\ 
{} & {} & \left(P_{out,SR}\right)^{N_S}P_{out,RD}\biggr],
\end{eqnarray}
where again the last term will decay much quicker than the others and can be ignored. The first term, after multiplication, will decay as 
$N_SN_D + N_SN_R$, while the second term decays as $N_SN_D + N_RN_D$.  Thus,
\begin{eqnarray}
\label{eqn:pout_lower}
P_{out} & \dot{\ge} & N_SN_D + N_R\min\{N_S,N_D\}.
\end{eqnarray}
Combining (\ref{eqn:pout_lower}) and (\ref{eqn:pout_upper}) we see that the proposed antenna selection achieves the full diversity gain in
the channel.

\section{Simulation Results}
We present a simple simulation to compare to a recent result on limited feedback beamforming~\cite{Khoshnevis2008}. 
For each case shown, we simulate the relay channel with
$N_S=N_R=N_D=3$ using BPSK modulation and an i.i.d. Rayleigh channel at each link. Bit error rate (BER) is the metric of interest. 
Figure~\ref{fig:versusKhosh} gives the results for $\mathbb{E}\{\gamma_{SR}^{(i)}\}=\mathbb{E}\{\gamma_{RD}^{(k)}\}=2$ dB for various
$\mathbb{E}\{\gamma_{SD}^{(i)}\}$. Note that this graph corresponds exactly to Fig. 9 in~\cite{Khoshnevis2008}, and we have included their
results for a Grassmannian codebook with more than 20 bits of feedback. Using antenna selection at both \sr{S} and \sr{R} requires 4 bits in 
this case and results in a loss of approximately 1 dB at high SNR. 

The theoretical lower bound of Figure~\ref{fig:versusKhosh} is when the source can simultaneously beamform the BPSK symbols
to both the relay and destination;
obviously this is an impossible task. The ``optimal'' performance curve was found numerically in~\cite{Khoshnevis2008} using gradient descent to
find a local optimum.

Figure~\ref{fig:diversity} shows the BER of uncoded BPSK versus $E_S/N_0$ for a relay channel with two antennas at each node. Note that increasing 
$E_S/N_0$ implies an increase in SNR at \emph{each link} (recall that noise terms are normalized and $\mathbb{E}|s|^2=E_S=\texttt{SNR}$).
The figure was generated using Monte Carlo simulations using $10^8$ channel realizations
for accuracy at high SNR, and demonstrates that antenna selection achieves the maximum diversity order available in the channel.

\section{Conclusion}
We explored antenna selection as a practical way of achieving the full diversity order of the nonregenerative MIMO relay channel. 
It was shown to achieve this diversity with a small SNR penalty relative to Grassmannian codebooks.

\bibliographystyle{IEEEtran}
\bibliography{IEEEabrv,MIMO_relay}

\begin{figure}[!b]
\centering
\includegraphics[width=3.5in]{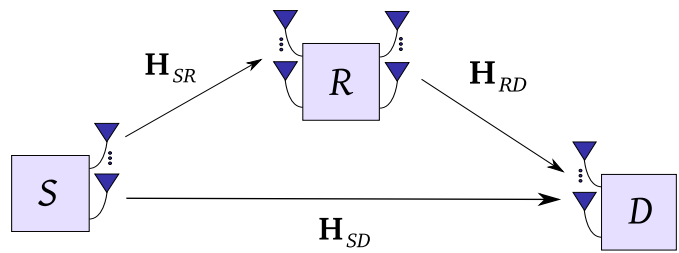}
\caption{The system model used in this letter. The source transmits in the first time slot, and the relay transmits in the second time slot.
The relay is shown with separate transmit and receive antennas for convenience; this assumption is not made in the analysis.}
\label{fig:MIMO_model}
\end{figure}

\begin{figure}[!b]
\centering
\includegraphics[width=3.5in]{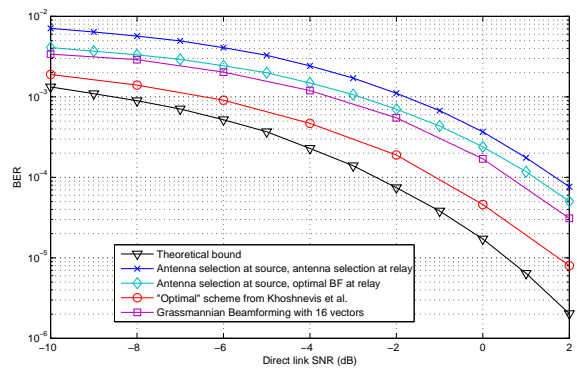}
\caption{BER performance for several MIMO amplify-and-forward beamforming strategies.}
\label{fig:versusKhosh}
\end{figure}

\begin{figure}[!b]
\centering
\includegraphics[width=3.5in]{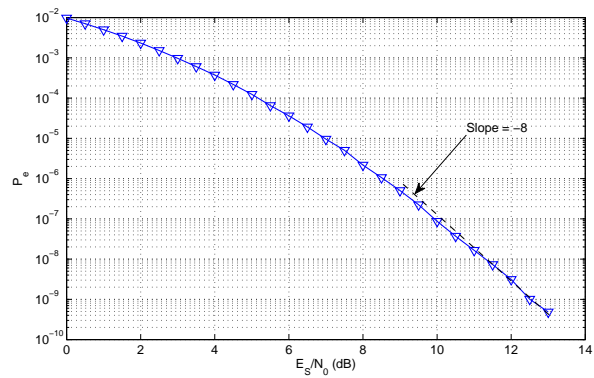}
\caption{BER performance for uncoded BPSK versus $E_S/N_0$ for i.i.d. Rayleigh fading. At high SNR, the slope of the curve approaches 
$-N_S(N_R+N_D)=-N_D(N_S+N_R)=-8$, which, as shown in Section IV, is the full diversity order of the channel.}
\label{fig:diversity}
\end{figure}

\end{document}